\documentstyle[12pt]{article}

\begin{document} 
\newtheorem{theorem}{Theorem}
\newtheorem{axiom}{Definition}
\newcommand{\nc}{\newcommand}

\nc{\theo}{\begin{theorem}}
\nc{\etheo}{\end{theorem}}
\nc{\bd}{\begin{axiom}}
\nc{\ed}{\end{axiom}}
\nc{\be}{\begin{equation}}
\nc{\ee}{\end{equation}}
\nc{\bea}{\begin{eqnarray}}
\nc{\eea}{\end{eqnarray}}
\nc{\no}{\nonumber \\}

\title{An Explanation of the Newman-Janis Algorithm}
\author{ S. P. Drake and Peter Szekeres \\ 
Department of Physics and Mathematical Physics, \\
University of Adelaide, Adelaide, \\ S. A. 5005, \\ Australia}
\date{\today}
\maketitle
\begin{abstract}
After the original discovery of the Kerr metric,
Newman and Janis showed that this solution could 
be ``derived'' by making an elementary complex 
transformation to the Schwarzschild solution.  
The same method was then used to obtain a new  
stationary axisymmetric  solution to Einstein's field equations
now known as the Kerr-newman metric,  
representing  a rotating massive charged black hole. 
However no clear reason has ever been given as to why the Newman-Janis 
algorithm works, many physicist considering it to be an {\it ad hoc} 
procedure or ``fluke'' and not worthy of further investigation. 
Contrary to this belief this paper shows why the 
Newman-Janis algorithm is successful in
obtaining the Kerr-Newman metric by removing some of the ambiguities present
in the original derivation. Finally we show that the only 
perfect fluid generated by the Newman-Janis 
algorithm is the (vacuum) Kerr metric and that the only Petrov typed D 
solution to 
the Einstein-Maxwell equations is the Kerr-Newman metric. 
\end{abstract}

\section{Introduction}
The generation of axisymmetric solutions to Einstein's field equations is a 
problem which has plagued physicists for many years. The unique class
of charged rotating black hole are described by Kerr-Newman spacetimes,
which can be generated by a technique known as the  
Newman-Janis algorithm~(NJA)~\cite{pap:nj1}~\cite{pap:nj2}. 
While this algorithm is particularly successful for ``deriving'' 
the Kerr metric and its electromagnetic generalization, the Kerr-Newman metric,
it has often been criticized~\cite{flah} on the grounds that (a) the procedure
is not a general method of generating vacuum from vacuum metrics, and (b)
there is a certain arbitrariness in the choice of complexification of terms
in the original seed metric (Schwarzschild or Reissner-Nordstr\"{o}m). 

Since the Kerr metric describes the spacetime of a rotating black hole, 
it is naturally of interest whether or not it  describes the 
exterior of an extended axisymmetric rotating massive body.
While Birkhoff's theorem~\cite{bk:mtw} tells us the spacetime exterior to a 
spherical symmetric massive body is uniquely described by the Schwarzschild
metric there is unfortunately no reason to expect that the 
spacetime exterior to an arbitrary stationary axisymmetric perfect fluid 
body should be the Kerr spacetime. 
In fact it remains an open question whether the Kerr solution can represent the exterior of any perfect fluid source at all.

The first genuine example of a Kerr interior solution appears to have been
provided by Neugebauer and Meinel~\cite{neug}, 
but it is rather difficult to access analytically, and as it represents a disc of matter the concept of perfect fluid can only apply in a degenerate sense. 
It would therefore be particularly attractive to generate interior Kerr
solutions by some simple procedure, and the obvious candidate appears
to be a generalization of the NJA, since that procedure is precisely
capable of generating the Kerr metric from the Schwarzschild metric.
The  possibility of such a generalization is given some encouragement
by Drake and Turolla's investigation~\cite{pap:dt}
of stationary axisymmetric metrics generated by the NJA
which match smoothly to the Kerr metric.
The original intention of this work was to generate new metrics which 
could be considered as perfect fluid sources of the Kerr metric
by application  of the NJA to static 
spherical symmetric seed metrics.
A class of metrics were found 
which match smoothly to the Kerr metric, and one particular test solution  
from this family was shown after detailed examination to be a ``physically sensible'' perfect fluid in the non-rotating limit.

The main purpose of this paper is (a) to specify precisely what the Newman-%
Janis algorithm is, (b) to understand under what circumstances and with what
choice of complexifications it will be successful in generating one solution
of Einstein's equations from another, and (c)  to explore the possibility of generalizing the algorithm to arbitrary seed metrics with the view to generating perfect fluid interior solutions of einstein's equations.

One key result from this paper is the demonstration that  
while the NJA is successful in generating metrics which match smoothly to 
the Kerr metric and are 
``physically sensible'' perfect fluids in the zero rotation limit these 
metrics are not rotating perfect fluids except in the vacuum case 
$(P = \rho = 0)$.  

In Section~\ref{sec:nja} we  describe the Newman-Janis Algorithm as a five-step procedure.  This may appear to be somewhat overspecific,  but these steps constitute the most general algorithm of this 
kind which has actually been found to work.
In Section~\ref{sec:enja} the algorithm is applied to
a general spherically symmetric spacetime.  Section~\ref{sec:pnja} presents the main
results, that non-vacuum perfect fluids can never be generated by the NJA,
while the Kerr-Newman solution is the most general algebraically special
spacetime which can be so generated. 
It follows as  a corollary that the Drake-Turolla metrics can never
provide a perfect fluid interior to the Kerr metric.
In Section~\ref{sec:con} the conclusion is drawn that the particular choice of complexification used in the standard NJA to generate the Kerr-Newman solution are not arbitrary, but could in fact be chosen in no other way in order for the NJA to be successful at all. 
This provides, in a sense, an ``explanation'' of the algorithm.

\section{The Newman-Janis algorithm} \label{sec:nja}
In this section we will describe the Newman-Janis algorithm in a form generalized from the original version used 
to generate the Kerr-Newman metric (rotating charged black hole) 
from the Reissner-Nordstr{\"o}m solution.
We treat the NJA as a five-step procedure for generating  new solutions
of Einstein's equations from known static spherically symmetric ones.  Whether
a similar process can apply to original seed metrics which are not
spherically symmetric is not known.

The five steps of the Newman-Janis  algorithm are as follows:   
\begin{enumerate}
\item Write a static spherically symmetric seed 
line element in  
advanced null coordinates $\{u,r,\theta,\phi\}$
\begin{equation} 
ds^2 = e^{2\Phi(r)}du^2 + e^{\Phi(r) + \lambda(r)}dudr -r^2
\left( d\theta^2
+ \sin^2\theta d\phi^2 \right). \label{eq:ds2}
\end{equation} 

In the Newman-Janis algorithm the seed was the Reissner-Nordstr{\"o}m
metric which in advanced Eddington-Finkelstein coordinates is 
\begin{equation} 
ds^2 = \left( 1 - \frac{2m}{r} - \frac{Q^2}{r^2} \right) du^2 + 2dudr - 
r^2\left( d\theta^2 + \sin^2\theta d\phi^2 \right).
\end{equation}

\item Express the contravariant form of the metric in terms of a 
null tetrad,
\begin{equation} 
g^{\mu\nu} = l^\mu n^\nu + l^\nu n^\mu-m^\mu\bar{m}^\nu 
- m^\nu\bar{m}^\mu, \label{eq:gmet}
\end{equation}
where
\[ l_\mu l^\mu = m_\mu m^\mu = n_\mu n^\mu = 0, \hspace{1em} 
l_\mu n^\mu = - m_\mu\bar{m}^\mu = 1, \hspace{1em} 
l_\mu m^\mu = n_\mu m^\mu = 0. \] 
For the spacetime~(\ref{eq:ds2}) the null tetrad vectors are
\begin{eqnarray}
l^\mu  & = & \delta^\mu_1 \nonumber 	\\
n^\mu & = & e^{-\lambda(r) - \Phi(r)}\delta^\mu_0 
- \frac12  e^{-2\lambda(r)} \delta^\mu_1 \nonumber \\
m^\mu & = & \frac{1}{\sqrt{2}r}
\left( \delta^\mu_2 
+ \frac{i}{\sin{\theta}}\delta^\mu_3 \right). \nonumber
\end{eqnarray}
It is also convenient to use the tetrad notation 
introduced by Newman and Penrose~\cite{pap:np} 
\[ Z_{a}^\mu = \left(l^\mu, n^\mu, m^\mu, \bar{m}^\mu \right), 
\hspace{2em} a = 1,2,3,4. \]

The null tetrad vectors for the Reissner-Nordstr{\"o}m metric are
\begin{eqnarray}
l^\mu  & = & \delta^\mu_1 	\nonumber\\
n^\mu & = & \delta^\mu_0 
- \frac12  \left( 1 - \frac{2m}{r} - \frac{Q^2}{r^2} \right) \delta^\mu_1 
\nonumber \\
m^\mu & = & \frac{1}{\sqrt{2}r}
\left( \delta^\mu_2 
+ \frac{i}{\sin{\theta}}\delta^\mu_3 \right). \nonumber
\end{eqnarray}

\item 
Extend the coordinates $x^\rho$ to a new set of complex coordinates
$\tilde{x}^\rho$ 
\[ x^\rho \rightarrow \tilde{x}^\rho = x^\rho + i y^\rho(x^\sigma), \]
where $y^\rho(x^\sigma)$ are analytic functions of the
real coordinates $x^\sigma$, and simultaneously let 
the null tetrad vectors $Z_a^\mu$ undergo a 
transformation 
\be
Z_a^\mu(x^\rho) \rightarrow 
\tilde{Z}_a^\mu(\tilde{x}^\rho, \bar{\tilde{x}}^\rho). \label{eq:tilde}
\ee

Finally we require that the transformation recovers the old tetrad and metric
when  $\tilde{x}^\rho = \bar{\tilde{x}}^\rho$. In summary the effect of
this ``tilde transformation'' is to create a new metric whose components
are (real) functions of complex variables,
\begin{equation}
g_{\mu\nu} \rightarrow \tilde{g}_{\mu\nu} : \tilde{\bf{x}} \times \tilde{\bf{x}} \mapsto {\rm I\! R},
\label{eq:cond1}
\end{equation}
while
\begin{equation}
\tilde{Z}_a^\mu(\tilde{x}^\rho,
\bar{\tilde{x}}^\rho)|_{\tilde{\bf{x}} = \bar{\tilde{{\bf{x}}}}} =
Z_a^\mu(x^\rho) \label{eq:cond2}
\end{equation}
The tilde transformation is clearly not unique as there are many different
choices  of the null tetrad vector coefficients which
satisfy the conditions~(\ref{eq:cond1})~and~(\ref{eq:cond2}).

In the original NJA, the tilde transformation on the
Reissner-Nordstr{\"o}m null tetrad vectors is
\begin{eqnarray}
l^\mu  & \rightarrow & \tilde{l}^\mu = \delta^\mu_1 	\label{eq:ntvl} \\
n^\mu  & \rightarrow & \tilde{n}^\mu = \delta^\mu_0 
- \frac12  \left( 1 - m\left(\frac{1}{\tilde{r}} + \frac{1}{\bar{\tilde{r}}}
\right)
 - \frac{Q^2}{\tilde{r}\bar{\tilde{r}}} \right) \delta^\mu_1 \label{eq:ntvn} \\
m^\mu & \rightarrow  & \tilde{m}^\mu = \frac{1}{\sqrt{2}\tilde{r}}
\left( \delta^\mu_2 
+ \frac{i}{\sin{\tilde{\theta}}}\delta^\mu_3 \right). \label{eq:ntvm}
\end{eqnarray}
A quick check shows that the above null tetrad vectors are those corresponding
to the Reissner-Nordstr{\"o}m metric when  $\tilde{x}^\rho = 
\bar{\tilde{x}}^\rho$.  However it is precisely here that a certain 
arbitrariness crept into the process, since the method of complexifying
the term $2m/r$ is quite different to the complexification of the 
$Q^2/r^2$ term.  It is our aim to provide some rationale for this
part of the NJ procedure.

\item  
A new metric is obtained by making a complex coordinate transformation
\begin{equation}
\tilde{x}^\rho  = x^\rho + i \gamma^\rho(x^\sigma) \label{eq:trans}
\end{equation}  
to  the  null tetrad vectors $\tilde{Z_a^\mu}$. The null tetrad 
vectors transform in the usual way
\[ Z^\mu_a = \tilde{Z}^\nu_a \frac{\partial x^\mu}
{\partial \tilde{x}^\nu}.  \]

The particular choice of complex transformations chosen by Newman and 
Janis to generate the Kerr-Newman metric were
\begin{equation} 
\tilde{x}^\rho = x^\rho + ia\cos{x^2}\left( \delta_0^\rho
 - \delta_1^\rho \right). \label{eq:trans1}
\end{equation}
From the transformed null tetrad vectors a new metric is recovered 
using~(\ref{eq:gmet}). For the null tetrad vectors given by 
Equations~(\ref{eq:ntvl}) to~(\ref{eq:ntvm})  
and the transformation given by~(\ref{eq:trans1}) the new metric 
with coordinates $x^\rho = \{ u, r, \theta, \phi\}$ in covariant 
form is 
\begin{equation} \label{eqn:coform}
g_{\mu\nu} =      \left(
\begin{array}{cccc}
1 - \frac{2mr - Q^2}{\Sigma} & 1 & 0 
& a\sin^2{\theta}\frac{2mr - Q^2}{\Sigma} \\
. & 0 & 0 & -a\sin^2{\theta} \\
. & . & -\Sigma & 0 \\
. & . & . & -\sin^2{\theta}\left( r^2 + a^2 - 
a^2\sin^2{\theta} \frac{2mr - Q^2}{\Sigma} \right) \\    
\end{array} \right). 
\end{equation}
As the metric is 
symmetric the ``.'' is used to indicate $g_{\mu\nu} = g_{\nu\mu}$. 
\[ \Sigma \equiv r^2 + a^2\cos{\theta}^2 \]. 
  
\item Finally it is assumed that a simple coordinate transformation of the 
form $u = t + F(r)$ , $\phi = \psi + G(r)$ will transform the metric to 
{\em Boyer-Lindquist coordinates}.
In this paper a set of  coordinates in which the 
metric has only one off-diagonal term $g_{t\phi}$ will be termed 
``Boyer-Lindquist''. 
  
To obtain the usual representation of the Kerr metric in 
Boyer-Lindquist coordinates, it is necessary to make a transformation 
on the null coordinate $u$ and the angle coordinate $\phi$
\begin{eqnarray}
u & = & t - \int \frac{a}{r^2 + a^2 + Q^2 - 2mr} dr \\
\phi & = & \psi - \int \frac{r^2 + a^2}{r^2 + a^2 + Q^2 - 2mr} dr
\end{eqnarray}
\end{enumerate}

\section{Extending the Newman-Janis algorithm} \label{sec:enja}
In the various stages of the NJA described above the only ambiguous point 
was the tilde transformation in step 3.
Applying this step to a general static spherically 
symmetric seed metric (\ref{eq:ds2}), 
the tilde operation produces the null tetrad vectors  
\begin{eqnarray}
\tilde{l}^\mu  & = & \delta^\mu_1 \nonumber	\\
\tilde{n}^\mu & = & e^{-\lambda(\tilde{r},\bar{\tilde{r}})-\phi(\tilde{r},
\bar{\tilde{r}})}\delta^\mu_0
-\frac12 e^{-2\lambda(\tilde{r},
\bar{\tilde{r}})}\delta^\mu_1 \nonumber \\
\tilde{m}^\mu & = & \frac{1}{\sqrt{2}\bar{\tilde{r}}}
\left( \delta^\mu_2 
+ \frac{i}{\sin{\tilde{\theta}}}\delta^\mu_3 \right). \nonumber
\end{eqnarray}
A new set of null tetrad vectors, and hence a new metric, results after the  
transformation~(\ref{eq:trans1}). These new null tetrad vectors are 
\begin{eqnarray}
l^\mu & = & \delta^\mu_1 \label{eq:ntvln} \\
n^\mu & = & e^{-\lambda(r,\theta) - \phi(r,\theta)}\delta^\mu_0
-\frac12 e^{-2\lambda(r,\theta)}\delta^\mu_1 
  \label{eq:ntvnn} \\
m^\mu & = & \frac1{\sqrt{2}(r + ia\cos{\theta})}\left(
ia\sin{\theta}(\delta^\mu_0 - \delta^\mu_1) + 
\delta^\mu_2 + \frac{i}{\sin{\theta}}\delta^\mu_3\right) \label{eq:ntvmn}
\end{eqnarray}
The coordinates $x^\rho = \{ u, r, \theta, \phi\}$ are all real.   
By equation~(\ref{eq:gmet}) the metric obtained from the null
tetrad vectors~(\ref{eq:ntvln}),~(\ref{eq:ntvnn}) and~(\ref{eq:ntvmn}) is,  
in covariant form, 
\begin{equation} \label{eq:coform}
g_{\mu\nu} =      \left(
\begin{array}{cccc}
e^{2\Phi(r,\theta)} & e^{\lambda(r,\theta) + \Phi(r,\theta)} & 0 & a\sin^2{\theta}
e^{\Phi(r,\theta)}(e^{\lambda(r,\theta)}- e^{\Phi(r,\theta)}) \\
. & 0 & 0 & -ae^{\Phi(r,\theta)+\lambda(r,\theta)}\sin^2{\theta} \\
. & . & -\Sigma & 0 \\
. & . & . & -\sin^2{\theta}(\Sigma + 
a^2\sin^2{\theta}e^{\Phi(r,\theta)}(2e^{\lambda(r,\theta)}-
e^{\Phi(r,\theta)})) \\    
\end{array}      \right)
\end{equation}
This completes steps 3 and 4, the generalization 
of the NJA without guessing the
tilde transformation~(\ref{eq:tilde}). At this stage the metric 
contains two unknown functions $\exp(\Phi)$ and $\exp(\lambda)$ of 
two variables $r,\theta$. The only constraints on these functions are 
given by~(\ref{eq:cond1}) and~(\ref{eq:cond2}). 
 
Step 5 is the transformation of the new metric into  Boyer-Lindquist 
coordinates by means of a transformation of the form 
$u = t +  \int g(r) dr$ and $\phi = \psi +  \int h(r) dr$ where the functions $g(r)$  and $f(r)$ will necessarily satisfy the equations
\begin{eqnarray}
g(r) & = & -\frac{e^{\lambda(r,\theta)}
(\Sigma + a^2\sin^2{\theta}e^{\lambda(r,\theta) + \Phi(r,\theta)})}
{e^{\Phi(r,\theta)}(\Sigma + a^2\sin^2{\theta}e^{2\lambda(r,\theta)})} 
\label{eq:g} \\
h(r) & = & -\frac{ae^{2\lambda(r,\theta)}}
{\Sigma + a^2\sin^2{\theta}e^{2\lambda(r,\theta)}}. \label{eq:h}
\end{eqnarray}
After some algebraic manipulations one finds that in these 
 coordinates $\{t,r,\theta,\psi \}$  
the metric is 
\begin{equation} 
g_{\mu\nu} =      \left(
\begin{array}{cccc}
e^{2\phi(r,\theta)} & 0 & 0 & a\sin^2{\theta}
e^{\phi(r,\theta)}(e^{\lambda(r,\theta)}- e^{\phi(r,\theta)}) \\
. & -\Sigma/(\Sigma e^{-2\lambda(r,\theta)} + a^2\sin^2{\theta})
 & 0 & 0 \\
. & . & -\Sigma & 0 \\
. & . & . & -\sin^2{\theta}(\Sigma + 
a^2\sin^2{\theta}e^{\phi(r,\theta)}(2e^{\lambda(r,\theta)}-
e^{\phi(r,\theta)})) \\    
\end{array}      \right). \label{eq:boy}
\end{equation}
By rearrangement of equation~(\ref{eq:h}) we find
\[e^{2\lambda(r,\theta)} = \frac{-h(r)\Sigma}{a^2 h(r) \sin^2\theta + a} =  \frac{\Sigma}{j(r) + a^2\cos^2\theta}. \]
 where $j(r) \equiv -a/h(r)- a^2$. 

In a similar manner  
equation~(\ref{eq:g}) may be used to express 
$\exp\Phi(r,\theta)$  in terms 
of the single variable functions $g(r)$ and $j(r)$,
\[ e^{\Phi(r,\theta)} =  \frac{\sqrt{\Sigma ( j(r) +  
 a^2\cos^2\theta)}}{k(r) + a^2\cos^2\theta}\] 
where $k(r) \equiv - g(r)(j(r) + a^2) - a^2$. 
The Boyer-Lindquist form of~(\ref{eq:coform}) is then
\begin{equation} \label{eq:blform2}
g_{\mu\nu} =      \left(
\begin{array}{cccc}
\frac{\Sigma\left( j(r) + a^2\chi^2 \right)}
{\left( k(r) + a^2\chi^2 \right)^2} & 0 & 0
& -\frac{a \left(1-\chi^2 \right)\left(j(r) - k(r)\right)
 \Sigma}{\left( k(r) 
+ a^2\chi^2\right)^2} \\
. & -\frac{\Sigma}{j(r) + a^2} & 0 & 0 \\
. & . & -\frac{\Sigma}{1-\chi^2} & 0 \\
. & . & . & -\left( 1 -\chi^2 \right)\Sigma\frac{\left( k(r) 
+ a^2 \right)^2 - a^2\left( 1 -\chi^2 \right) \left( j(r) + a^2 \right)}
{\left(k(r) +a^2\chi^2 \right)^2} \\
\end{array}      \right). 
\end{equation}
Where $\chi \equiv \cos\theta$ so that 
\[ \Sigma \equiv r^2 + a^2\chi^2. \]

In order to calculate properties of the metric tensor~(\ref{eq:blform2}) 
the packages {\it Tensor} and {\it Debever} where used inside {\it Maple V}.
It is a well known phenomena that while humans often prefer 
to work with trigonometric  functions computers do not. 
The cause of this problem is that there is not a unique way to 
simplifying trigonometric  functions. The safest way to remove  
this problem is 
to avoid using trigonometric functions altogether in computer aided 
calculations. For this reason the substitution $\chi \equiv \cos\theta$
was made. 

\section{Properties of metrics generated by 
the Newman-Janis algorithm} \label{sec:pnja}
The package {\it Tensor} in {\it Maple V} allows us to calculate the 
Einstein tensor any metric tensor. {\it Debever} calculates the Newman-Penrose
spin coefficients.
Below, we provide some theorems for these spacetimes using the above mentioned
packages.  The algebraic expressions of the Einstein tensor 
and the spin coefficients tend to be rather 
lengthy, fortunately all those of interest to 
us can be expressed in the form $\sum_m H_m \chi^{2m}$ where $H_m$ is
a function of $r$ only. For reasons of  compactification all the
curvature expressions will be written in this way and the specific 
forms for $H_m$ will be shown only when required. 
The interested reader is encouraged to check these expressions of {\it Maple},
{\it Mathematica} or their favorite algebraic manipulation program. 
\begin{theorem} The only perfect fluid generated by the Newman-Janis 
Algorithm is the vacuum.
\end{theorem}
{\bf Proof: } 
The Einstein tensor resulting from the metric~(\ref{eq:blform2}) has two off
diagonal terms, $G_{t\Phi}$ and $G_{r\chi}$. If the Einstein tensor is 
equivalent to the stress energy tensor of a perfect fluid
\begin{equation}
G_{\mu\nu} = (P + \rho)U_\mu U_\nu  - P g_{\mu\nu} \label{eq:eten}
\end{equation}
then it is required that $G_{r\chi} = 0$. The reason for this is that  
the four velocity $U_\nu$ being a time-like vector must have $U_0 \neq 0$. 
Generation of the Einstein tensor reveals $G_{tA} = 0$ if $A = r$ 
or $A= \chi$. From equations~(\ref{eq:blform2}) and~(\ref{eq:eten}) 
it follows that $U_{r} = U_{\chi} = 0$. Since $g_{t\phi}$ is the only non-zero
component of $g_{\mu\nu}$ it follows that
$G_{r\chi}$ must vanish identically.  
The $G_{r\chi}$ component of the Einstein tensor generated by {\it Tensor} 
in {\it Maple V} is 
\[
-3\chi a^2 \frac{\left(2r- k(r)'\right)a^4\chi^4 + \left( 2k(r) 
- k(r)'r\right)2ar\chi^2 - k(r)'r^4 + 2rk(r)^2}
{\Sigma^2\left(k(r) +a^2\chi^2\right)^2},
\]  
where `` $'$ '' denotes the derivative with respect to $r$. This expression 
vanishes if and only if 
\begin{equation}
k(r) = r^2. \label{eq:k1}
\end{equation}
To resolve $j(r)$ with this definition of $k(r)$ we look at the isotropic
pressure condition, 
\bea
G_{rr}/g_{rr} & - & G_{\chi\chi}/g_{\chi\chi}  =  0 \no 
\frac{-1}2  \frac{(j(r)'' -2)a^2\chi^2}{r^2 + a^2\chi^2} & - & 
\frac12\frac{2r^2  + r^2 j(r)'' -4 r j(r)'
+ 4 j(r)}{r^2 + a^2\chi^2}   =  0. \label{eq:iso} 
\eea
As $\chi$ is an independant variable the isotropic pressure 
condition~(\ref{eq:iso}) is satisfied if and only if 
\bea 
j(r)'' & - & 2  =  0 \no
j(r)'' -2)a^2\chi^2 + 2r^2  +  r^2 j(r)'' & - & 4 r j(r)'
+ 4 j(r)  =  0 \nonumber.
\eea
The unique solution to this pair of equations is 
\begin{equation}
j(r) = r^2 + d_1 r, \label{eq:j1}
\end{equation}
where $d_1$ is a  constant of integration. 

Substituting equations~(\ref{eq:j1})
and~(\ref{eq:k1}) into the metric 
(\ref{eq:blform2}) generated by the NJA, and 
setting the constant of 
integration $d_1$ to equal twice the mass we get the Kerr metric in 
Boyer-Lindquist coordinates. $\hspace{2cm}\Box$
\begin{theorem} The only algebraically special spacetimes generated by 
the Newman-Janis algorithm are Petrov type D. 
\end{theorem}
{\bf Proof: }
It was shown in section~\ref{sec:enja} that since the metric~(\ref{eq:coform}) 
can be transformed to Boyer-Lindquist 
coordinates the functions $\exp\Phi(r,\theta)$ and $\exp\lambda(r,\theta)$
can be expressed in the form
\begin{eqnarray}
e^{\lambda(r,\chi)} & = & \frac{\sqrt{\Sigma}}
{\sqrt{j(r) + a^2\chi^2}} \label{eq:lam} \\
e^{\Phi(r,\chi)} & =  & \frac{\sqrt{\Sigma\left( j(r) + 
a^2\chi^2 \right)}}{k(r) 
+ a^2\chi^2} \\
\Sigma & \equiv  & r^2 + a^2 \chi^2 \\
\chi & \equiv  & \cos\theta. \label{eq:chi2}
\end{eqnarray}
The resulting null tetrad vectors are
\begin{eqnarray}
l^\mu & = & \delta^\mu_1 \label{eq:blntvl} \\
n^\mu & = & \frac{k(r) + a^2\chi^2}{\Sigma} \delta^\mu_1
-\frac12 \frac{j(r) + a^2\chi^2}{\Sigma}\delta^\mu_0 \label{eq:blntvn} \\ 
m^\mu & = & \frac1{\sqrt{2}(r + ia\chi)}\left(
ia\sqrt{1-\chi^2} (\delta^\mu_0 - \delta^\mu_1) + 
\delta^\mu_2 + \frac{i}{\sqrt{1-\chi^2}}\delta^\mu_3\right) \label{eq:blntvm}
\end{eqnarray}
Using the package {\it Debever} in {\it Maple V} it is possible to 
compute the Newman-Penrose coefficients~\cite{pap:np} from 
the null tetrad vectors~(\ref{eq:blntvl}),
(\ref{eq:blntvn}) and (\ref{eq:blntvm}).
It is found that $\Psi_0$ is identically zero. 
A spacetime is said to be algebraically  
special~\cite{bk:hi} if $\Psi_0 = \Psi_1 = 0$. 
$\Psi_1$  is a rather long expression which can however
be expressed as 
\be
\Psi_1 = \frac{\sum_{m=0}^4 i K_m \chi^{2m}}{\Sigma^2(r-ia\chi)
(k(r)+a^2\chi^2)\sqrt{1-\chi^2}}, \label{eq:psi1}
\ee
where $K_m$ are functions of $k(r)$ only. That is, 
the vanishing of $\Psi_1$ is independent of $j(r)$. As $r$ and $\chi$ are
independent variables, $\Psi_1 = 0$ if and only if $K_m =0$ for 
$m= 0,1,2,3,4$. 
\[K_4 = k''(r) - 2 \]
 which equals zero if and only if  
$k(r) = r^2 + c_1 r + c_0$. 
Substituting this expression for $k(r)$ into $K_3$  
it is found that $K_3 = a^4(4c_0 -c1^2)$ which equals zero for $a \neq 0$
if and only if $c_1^2 - 4c_0 =0$  
so that 
\begin{equation}
k(r) = r^2 + c_1(r+c_1/4). \label{eq:k2}
\end{equation}
Furthermore with $k(r)$ given by equation~(\ref{eq:k2}) it is 
easy to show by direct substitution that $K_m = 0$ 
for all allowed values of $m$. 
 
Hence all spacetimes generated by the Newman-Janis Algorithm which are 
algebraically special uniquely satisfy equation~(\ref{eq:k2}).
The proof that they also Petrov type D involves substituting the 
expression for $k(r)$ into the expressions for 
the Newman-Penrose spin coefficient $\Psi_i$ and 
showing that they satisfy the relation
\[
\Psi_2\Psi_4 - 2\Psi_3^2/3 = 0. 
\]
This can be checked  checked with the {\it Debever} package in {\it Maple V}.
$\hspace{2cm}\Box$
\begin{theorem}
The only Petrov type D spacetime generated by the Newman-Janis algorithm 
with a vanishing Ricci scalar is the Kerr-Newman spacetime.  
\end{theorem} 
{\bf Proof: }
Solutions to the 
Einstein-Maxwell field equations have a vanishing Ricci scalar. 
Once again using {\it Tensor} within {\it Maple V} 
it is possible to calculate  
the Ricci scalar from the metric~(\ref{eq:blform2}) with the 
definition~(\ref{eq:k2}). By grouping the Ricci scalar $R$ into powers of 
$\chi$ 
\[
R = \frac{\sum_{m=0}^4 J_m \chi^{2m}}{\Sigma^3\left( r^2 + c_1(r + 
c_1/4)\right)^2},
\]
$J_m$ are functions of $j(r)$ only.
As $r$ and $\chi$ are
independent variables $R  = 0$ if and only if $J_m =0$ for all allowed 
values of $m$. $J_4 = j(r)'' - 2$   
which equals zero if and only if $j(r) = r^2 + d_1 r + d_0$.
Substituting this expression for $j(r)$ along with~(\ref{eq:k1})  
into $R$ it is found that 
\[ R = c_1 \frac{ \sum_{m=0}^3 I_m \chi^{2m}}{\Sigma^3\left( r^2 + c_1(r + 
c_1/4)\right)^2}, \]
where $I_m$ depends on the constants $c_1, d_1,d_0$ and the variable $r$.   
It is not difficult, though tiresome, to show that 
if you do not assume that $c_1 =0$ then $c_1 = -2 d_1$ and
$d_1 = d_0 = 0$. 
The vanishing of $R$ is assured if and only if 
$c_1 = 0 $.  In which case the functions $j(r)$ and $k(r)$ are
\begin{eqnarray}
j(r) & = & r^2 + d_1 r + d_0 \nonumber \\
k(r) & = & r^2 \nonumber
\end{eqnarray}
Setting the constants of integration $d_1$ and $d_0$ to be twice the mass and
the square of the charge of the black hole respectively one obtains the 
Kerr-Newman metric. $\hspace{3cm}\Box$

\section{Conclusion} \label{sec:con}

Previously all work on the 
Newman-Janis algorithm has involved some  guess work. It was noticed that 
the Kerr-Newman metric could be obtained if a complex extension to the 
metric coefficients of the 
Reissner-Nordstr\"{o}m  seed metric was made 
\[ ds^2 = \left( 1 - m\left( \frac{1}{r} + \frac{1}{\bar{r}} \right) 
+ \frac{Q^2}{r\bar{r}} \right) dt^2 
- \left( \frac{r\bar{r}}{r\bar{r} - m (r + \bar{r}) + Q^2} \right) dr^2
- r\bar{r} d^2 \theta - r\bar{r} \sin^2{\theta} d^2\phi \] 
before applying the NJA,  where the bar denotes the complex conjugate 
of a particular variable. The only reason given for doing this 
was that it was successful. Our analysis does not rely on any such guess work 
and gives an  unambiguous 
explanation of the success of the NJA in generating  Kerr-Newman  metric.  

In this paper we have proved the following:
1) The only perfect fluid spacetime generated by applying the 
Newman-Janis algorithm to a static spherically symmetric seed metric
which may be written in Boyer-Lindquist form is the Kerr metric. 
2) The only algebraically special spacetimes  generated by applying the 
Newman-Janis algorithm to  static spherically symmetric seed metrics 
which may be written in Boyer-Lindquist form are Petrov-type D.  
3) The only algebraically 
special spacetime generated by applying the Newman-Janis algorithm to 
a static spherically symmetric seed metric
which may be written in Boyer-Lindquist form and which has vanishing Ricci
scalar (e.g.\ is a solution of the Einstein-Maxwell equations)
is the Kerr-Newman metric. 

The relation of this work to the previous results 
of Drake and Turolla~\cite{pap:dt} is  that while the NJA is 
successful in generating interior spacetimes which match smoothly to the 
Kerr metric, even if these interiors are perfect fluids in the non rotating 
limit this is not the case when rotation is included. 
\section*{Acknowledgments} 
All of algebraic calculations presented in this paper were done
with the aid of {\it Maple V} and  using the routines  {\it Tensor} 
and {\it Debever} which are contained therein. 
SPD would like to thank the Australian Post-graduate Programme 
for support during the completion of this work.
SPD would 
also like to thank D. Hartley,
M. Howes, T. Rainsford, R. Turolla and D. Wiltshire for many helpful 
discussions.


\begin{thebibliography}{99}

\bibitem{pap:nj1} E. T. Newman and A. I. Janis, {\it J. Math. Phys.} {\bf 6}
915 (1965).
\bibitem{pap:nj2} E. T. Newman, E. Couch, K. Chinnapared, A. Exton, A. 
Prakash and R. Torrence, {\it J. Math. Phys.} {\bf 6} 918 (1965).
\bibitem{flah} E.J. Flaherty {\it Hermitian and K\"{a}hlerian geometry in
relativity}, Lecture notes in Physics {\bf 46} (Springer-Verlag, Berlin 976).
\bibitem{bk:mtw} C. W. Misner, K. S. Thorne and J. A. Wheeler, 
{\it Gravitation} (W.H. Freeman and Co., San Francisco, 1973).
\bibitem{neug} G. Neugebauer and R. Meinel, {\it Phys. Rev. Lett.}
{\bf 73}, 2166 (1994)
\bibitem{pap:dt} S. P. Drake and R. Turolla, {\it Class. Quantum Grav. }
{\bf 14} 1883 (1997).
\bibitem{pap:np} E. T. Newman and R. Penrose, {\it J. Math. Phys.} 
{\bf 3} 566 (1962).
\bibitem{bk:hi} S. W. Hawking and W. Israel, {\it General Relativity:
An Einstein Century Survey},   (Cambridge University Press, 
Cambridge, 1979).
\end{thebibliography}
\end{document}